\documentclass[a4paper,referee]{raa}            

\usepackage{graphicx,times}  

\usepackage{natbib}
\usepackage{booktabs}
\usepackage{amssymb,amsmath}
\bibpunct{(}{)}{;}{a}{}{,}
\usepackage{lineno}
\usepackage{CJK}
\usepackage[pagebackref=false,colorlinks,linkcolor=blue,anchorcolor=blue,citecolor=blue]{hyperref}

\begin{document}

   \title{The Mini-SiTian Array: A Pathfinder for the SiTian Project}

   \volnopage{Vol.0 (20xx) No.0, 000--000}      
   \setcounter{page}{1}          

   \author{{Yang Huang}\inst{1,2}
    \and {Jifeng Liu}\inst{1,2}
    \and {Hong Wu}\inst{1}
     \and {Zhaohui Shang}\inst{1}
      \and {Ali Luo}\inst{1}
       \and {Shaoming Hu}\inst{3}
       \and {Wenyuan Cui}\inst{4}
    \and {Yongna Mao}\inst{1,2}}
   \institute{National Astronomical Observatories, Chinese Academy of Sciences, Beijing 100101, China; \\
    \and
    School of Astronomy and Space Science, University of Chinese Academy of Sciences, Beijing 100049, China;\\
    \and
    Shandong Key Laboratory of Optical Astronomy and Solar-Terrestrial Environment, School of Space Science and Technology, Institute of Space Sciences, Shandong University, Weihai, Shandong, 264209, China;\\
    \and
    Department of Physics, Hebei Normal University, Shijiazhuang 050024, China\\
\vs\no
   {\small Received~~20xx month day; accepted~~20xx~~month day}}

\abstract{
The Mini-SiTian Array serves as a pathfinder for the SiTian project, which aims to survey the entire sky in $gri$ bands every 30 minutes, reaching a limiting magnitude of 21. This special issue features 11 papers covering the design, operation, data reduction, and early scientific results from two years of Mini-SiTian observations. The insights gained from these pathfinder experiments represent a significant milestone toward the full realization of the SiTian project.
}

   \authorrunning{Huang et al.}            
   \titlerunning{The Mini-SiTian Array}  

   \maketitle

\section{SiTian project}
Astronomy is transitioning into an era driven by large-scale time-domain surveys, offering unprecedented opportunities to monitor the dynamic universe. International projects such as the Zwicky Transient Facility \cite[ZTF;][]{M19}, and domestic initiatives like the Wide Field Survey Telescope \citep[WFST;][]{WTG23} and Multi-channel Photometric Survey telescope \citep[Mephisto;][]{Y20}, are propelling time-domain observations into a golden age. However, current all-sky monitoring remains limited in cadence, with even future flagship surveys, such as the Large Synoptic Survey Telescope (LSST; \citealt{I19}), revisiting the entire southern sky only once every three days.

Many high-energy astrophysical transients, including supernovae, tidal disruption events (TDEs), and the electromagnetic counterparts of gravitational wave events, exhibit their most critical early-time evolution within a few hours or less. Capturing these rapid changes requires an observational strategy that enables continuous, high-cadence all-sky monitoring. To address this challenge, the National Astronomical Observatories of China (NAOC) has proposed a global array of at least 60 meter-class wide-field telescopes, called the SiTian project, to achieve full sky coverage. This initiative aims to perform a three-band ($g, r, i$) survey reaching a depth of 21 mag with a cadence of 30 minutes \citep{LJF21}.

\section{The Mini-SiTian Array}
To lay the foundation for this ambitious project, the SiTian collaboration team has developed and deployed a prototype telescope array, known as the Mini-SiTian (MST) array, which consists of three 30 cm telescopes. The MST array is located at Xinglong Observatory and includes the three telescopes: MST1, MST2, and MST3. All telescopes feature a refractive optical configuration, each with a 300 mm primary mirror and a focal ratio of f/3. The optical design is detailed in \cite{H25a}.
Each telescope is equipped with a camera positioned at the Cassegrain focus. The chosen camera is the ZWO ASI6200 CMOS, which has a resolution of $9,576 \times 6,388$ pixels, a pixel size of $3.76\,\mu\text{m}$, and a pixel scale of $0.862''/\text{pixel}$. This configuration provides a field of view (FOV) of $2.29^\circ \times 1.53^\circ$.
The performance of the three cameras was thoroughly evaluated by \cite{Z25} under standard laboratory conditions, following established protocols for astronomical camera testing. A key finding was that these commercial cameras exhibit a linearity accuracy of 0.3\%, meeting the high-precision photometry requirements and performing on par with state-of-the-art CCD detectors.
To ensure seamless operation of the telescopes, a master control system was developed by \cite{W25a}. The MST survey began in November 2023, and its first two years of operation were summarized by \cite{He25}, covering the sky survey strategy, field selection, and operational statistics.

\section{Data Processing Pipeline of the MST Array}
The MST team has developed an advanced data processing pipeline capable of handling the generated data in both static and real-time modes. The entire processing workflow is comprehensively outlined by \cite{X25a}. The static pipeline includes the removal of instrumental effects, as well as standard astrometry and photometry.
A rigorous calibration procedure was implemented for photometry, achieving a CMOS photometric precision of 5 mmag for bright sources ($G_{\rm MST} \sim 13.5$ mag) with a typical exposure time of 300 seconds \citep{X25b}.
Such high precision, better than 1\%, is achieved for the first time in a CMOS-detector-based survey and is comparable to that of surveys using CCD cameras.
To detect transients and rare variable stars, the team developed a real-time pipeline, described by \cite{G25}. This pipeline is capable of generating differential images and detecting sources within 150 seconds, which is sufficient given the typical 5-minute exposure time of MST observations.
Additionally, a neural network classifier was developed by the team to distinguish genuine signals from false positives for the identified variable sources or transients \citep{S25}.

\section{Early Science Results of the MST Array}

Throughout its two-year survey, the MST team has defined key scientific goals, target selection criteria, and observational strategies \citep{H25b}.
Using two years of observations from one selected field (``F02" in the survey), \cite{Liu25} obtained light curve measurements for 22 asteroids, 14 of which were newly identified.
Additionally, \cite{W25b} simulated the expected detection rate of tidal disruption events (TDEs) by the MST Array.
As gravitational wave electromagnetic counterparts, such as kilonovae, are a key scientific objective of the SiTian project, \cite{L25} simulated potential kilonova detections during the LIGO O4 observing run using the SiTian prototype telescope.

Over the two years of Mini-SiTian array operations, we have built a highly capable team with expertise in hardware testing, telescope operation and maintenance, data processing, and time-domain astronomy. This experience has provided invaluable technical and operational insights, forming a strong foundation for the successful deployment of the full-scale SiTian project.

\begin{acknowledgements}
Y.H. acknowledges the supports from the National Key Basic R\&D Program of China via 2023YFA1608303, the Strategic Priority Research Program of the Chinese Academy of Sciences (XDB0550103) and the National Science Foundation of China (NSFC) through grant Nos. of 12422303 and 12261141690.
J.F.L. acknowledges support the NSFC through grant Nos. of 11988101 and 11933004, and support from the New Cornerstone Science Foundation through the New Cornerstone Investigator Program and the XPLORER PRIZE.
H.W. acknowledges support from the NSFC through grant Nos. 12090041 and 12090040.
S.M.H. acknowledges support from the NSFC through grant No. 11873035.
W.Y.C. acknowledges support from the NSFC through grant No. 12173013.

The SiTian project is a next-generation, large-scale time-domain survey designed to build an array of over 60 optical telescopes, primarily located at observatory sites in China. This array will enable single-exposure observations of the entire northern hemisphere night sky with a cadence of only 30-minute, capturing true color (gri) time-series data down to about 21 mag. This project is proposed and led by the National Astronomical Observatories, Chinese Academy of Sciences (NAOC). As the pathfinder for the SiTian project, the Mini-SiTian project utilizes an array of three 30 cm telescopes to simulate a single node of the full SiTian array. The Mini-SiTian has begun its survey since November 2022. The SiTian and Mini-SiTian have been supported from the Strategic Pioneer Program of the Astronomy Large-Scale Scientific Facility, Chinese Academy of Sciences and the Science and Education Integration Funding of University of Chinese Academy of Sciences.
\end{acknowledgements}

\label{lastpage}


\begin{thebibliography}{}
\bibitem[Gu et al. (2025)]{G25} Gu, H.-R., Huang, Y., Sun, Y.-K. et al., 2025, RAA, accepted
\bibitem[Han et al. (2025a)]{H25a} Han, Z.-J., Li, Z.-Y., Chen, C. et al., 2025a, RAA, accepted
\bibitem[Han et al. (2025b)]{H25b} Han, H.-G., Huang, Y., Wang, B.-C. et al., 2025b, RAA, accepted
\bibitem[He et al. (2025)]{He25} He, M., Wu, H., Ge, L. et al., 2025, RAA, accepted
\bibitem[Ivezi{\'c} et al.(2019)]{I19} Ivezi{\'c}, {\v{Z}}., Kahn, S.~M., Tyson, J.~A., et al.\ 2019, \apj, 873, 111. doi:10.3847/1538-4357/ab042c
\bibitem[Li et al. (2025)]{L25} Li, Z.-R., Gu, H.-R., Sun, Y.-K. et al., 2025, RAA, accepted
\bibitem[Liu et al. (2025)]{Liu25} Liu, Z.-X., Gao, J., Gu, H.-R. et al., 2025, RAA, accepted
\bibitem[Liu et al.(2021)]{LJF21} Liu, J., Soria, R., Wu, X.-F., et al.\ 2021, Anais da Academia Brasileira de Ciencias, 93, 20200628. doi:10.1590/0001-3765202120200628
\bibitem[Masci et al.(2019)]{M19} Masci, F.~J., Laher, R.~R., Rusholme, B., et al.\ 2019, \pasp, 131, 018003. doi:10.1088/1538-3873/aae8ac

\bibitem[Shi et al. (2025)]{S25} Shi, J.-H., Gu, H.-R., Huang, Y. et al., 2025, RAA, accepted
\bibitem[Wang et al. (2025a)]{W25a} Wang, Z., Zou, J.-H., Hu, Y. et al., 2025a, RAA, accepted
\bibitem[Wang et al. (2025b)]{W25b} Wang, B.-C., Jin, J.-J., Zhang, Y. et al., 2025b, RAA, accepted
\bibitem[Wang et al.(2023)]{WTG23} Wang, T., Liu, G., Cai, Z., et al.\ 2023, Science China Physics, Mechanics, and Astronomy, 66, 109512. doi:10.1007/s11433-023-2197-5
\bibitem[Xiao et al. (2025a)]{X25a} Xiao, K., Li, Z.-R., Huang, Y. et al., 2025a, RAA, accepted
\bibitem[Xiao et al.(2025b)]{X25b} Xiao, K., Huang, Y., Yuan, H., et al.\ 2025b, \apjl, 982, L27. doi:10.3847/2041-8213/adbd3c
\bibitem[Yuan et al.(2020)]{Y20} Yuan, X., Li, Z., Liu, X., et al.\ 2020, \procspie, 11445, 114457M. doi:10.1117/12.2562334
\bibitem[Zhang et al. (2025)]{Z25} Zhang, Y., Du, L., Hu, Y. et al., 2025, RAA, accepted
\end{thebibliography}
\end{document}